# The hippocampal-striatal circuit for goal-directed and habitual choice

Fabian Chersi
Institute of Cognitive Neuroscience, UCL, London, UK
f.chersi@ucl.ac.uk

**Abstract**

It is now widely accepted that one of the roles of the hippocampus is to maintain episodic spatial representations, while parallel striatal pathways contribute to both declarative and procedural value computations by encoding different input-specific outcome predictions.

In this paper we investigate the use of these brain mechanisms for action selection, linking them to model-based and model-free controllers for decision making. To this aim we propose a biologically inspired computational model that embodies these theories and explains the functioning of the hippocampal-striatal circuit in a rat navigation task. Its main characteristic is to allow the cooperation of habitual and goal-directed behaviors, with the hippocampus primarily involved in encoding spatial information and simulating possible navigation paths, and the ventral and dorsal striatum involved in learning stimulus-response behaviors and evaluating the reward expectancies associated to predicted locations and sensed stimuli, respectively.

The architecture we present employs an unsupervised reinforcement learning rule for the hippocampal-striatal network that is able to build a representation of the environment in which rewarding sites and informative landmarks produce value gradients that are used for planning and decision making. Additionally, it utilizes an arbitration mechanism that balances between exploitation, i.e. stimulus-response behaviors, and mental exploration, i.e. motor imagery processes, based on the intensity and the variability of the responses of striatal neurons.

We interpret these results in light of recent experimental data that show anticipatory activations in hippocampal and striatal areas.

**Introduction**

Traditionally, memory is considered to be of two types: declarative and procedural. The first one refers to the capacity to deliberatively and consciously recall events and situations from the past. The second one indicates the capabilities to automatically and unconsciously execute motor or cognitive skills.

Experimental findings indicate that the Hippocampus (HPC) is strongly involved in declarative memory, storing information about experienced events set in a specific spatio-temporal context (Eichenbaum et al., 2007; Tulving and Markowitsch, 1998).

The most striking characteristic of this area is the presence of so called *place cells* (O'Keefe and Nadel, 1978; Redish, 1999), neurons that possess high spatial selectivity, primarily firing within a small area (the place field) of the rat's environment. Interestingly, place cell activity may occur outside a cell's place field, giving origin to "forward sweeps" (Johnson and Redish, 2007) or memory "replays" (Foster and Wilson, 2006; Jackson et al., 2006), signaling an internally driven activation of these cells possibly for the purpose of trajectory planning or evaluation of potential alternatives (Chersi et al., 2013; Daw et al., 2005; Jensen and Lisman, 2005).

On the other hand, the Striatum and the connected Basal Ganglia (BG) structures are known to come into play during procedural learning and habit formation through the modifications in the cortico-basal ganglia loops (Chersi, 2012; McDonald and White, 1993; Packard and McGaugh, 1992; Yin and Knowlton, 2006).

The Striatum is subdivided into three regions: the dorsolateral striatum (DLS), which mediates stimulus–response learning and habit formation, the dorsomedial striatum (DMS), which is associated with cognitive functions and action–outcome learning, and ventral striatum (VS), which is involved in motivational and affective processing (Packard and McGaugh, 1992; Voorn et al., 2004). Interestingly, the VS eludes the strict classification introduced above as it also receives strong projections from the Hippocampus (Groenewegen et al., 1987).

In the following paragraphs we will start by describing the experimental setup used to validate the model, then we describe in detail the architecture at the circuit level and then the mathematical equations that govern the behavior and the connectivity of neurons.

Finally, we report simulation results showing the decision making and learning capabilities of the system and elaborate on how the different areas interact and what role they play contribute during spatial navigation.

**Experimental setup**

In our setup, a simulated rat can freely explore a complex 2-D maze (see Figure 1) of the size of 2.25x2.40 meters, which contains a home position (lower left corner), where the rat is repositioned at the beginning of every new trial, two locations with different types of rewards: food (cheese) and water, respectively, and one landmark. Paths are delimited by walls that contain colored patches. These have been added in order to work as visual cues for self-localization and navigation.

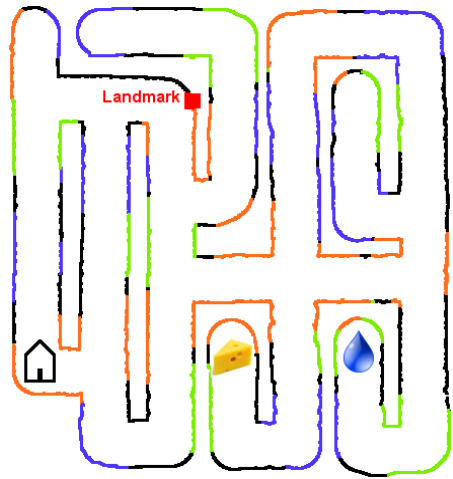

**Figure 1**. Representation of the maze utilized in these simulations with its most important features: in the bottom left corner is the Home position where the rat is replaced after each successful completion of the task. In the lower right part there are the two types of reward: cheese and water. Color patches have been added to the walls in order to improve self-localization.

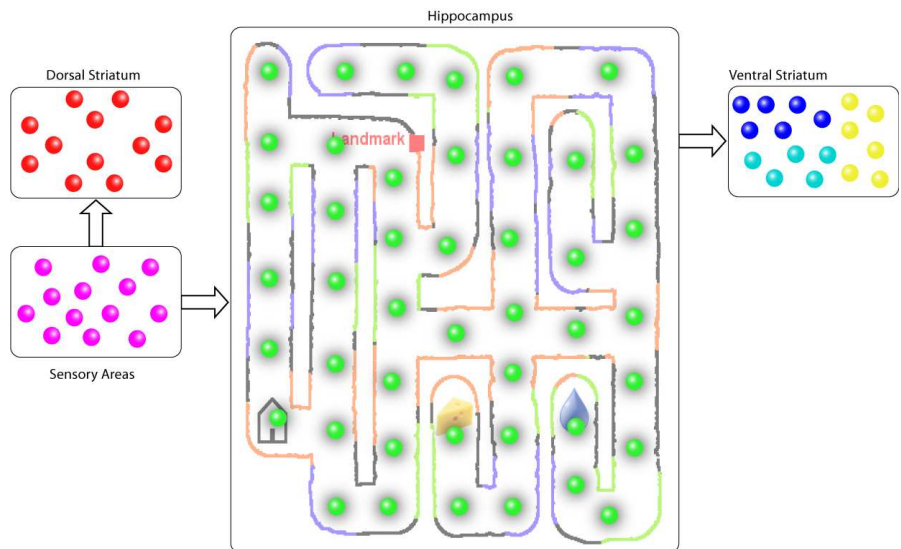

**Figure 2.** Representation of the complete neural architecture overlaid to the maze. Each place cell in the Hippocampus (green spheres) is receptive for a small portion of the maze and projects to the value neurons in the Ventral Striatum thus allowing for a conjunct representation of the value of locations. Visual neurons (purple spheres) encode the length of the line of sight and the color of the intersected object. Sensory patterns are used to learn stimulus-response associations through the Dorsal Striatum.

**Model architecture**

The brain circuit described above (see Figure 2) has been implemented as a modular network of firing rate based neurons.

The first module represents the visual areas and has been implemented as a mupli-layer perceptron with 121x4=484 neurons that encode the "colored" range map. More precisely, each of the 121 "lines of sight" (see Figure 3) that spawn the interval between -120º and +120º (referred to the head direction, every 2º) is encoded by 4 neurons, which represent the distance to the observed object, and its color in RGB format, respectively. This sensory vector may be interpreted as a combination of a mono-dimensional depth map and a mono-dimensional image, both 121 wide and 1 pixel high. The (sensory) input to this network has the following form:

$$p = (d_1, r_1, g_1, b_1, d_2, r_2, g_2, b_2, ..., d_n, r_n, g_n, b_n) \qquad (1)$$

where $d_n$ is the length of the *n*-th line of sight (*n*=1,… ,121), and $r_n, g_n, b_n$ define the color of the object seen in that specific direction.
The hidden layer contains 500 neurons, while the output layer contains 200 neurons.
The function of this module is to classify sensory patterns for the use of the hippocampus and the dorsal striatum.

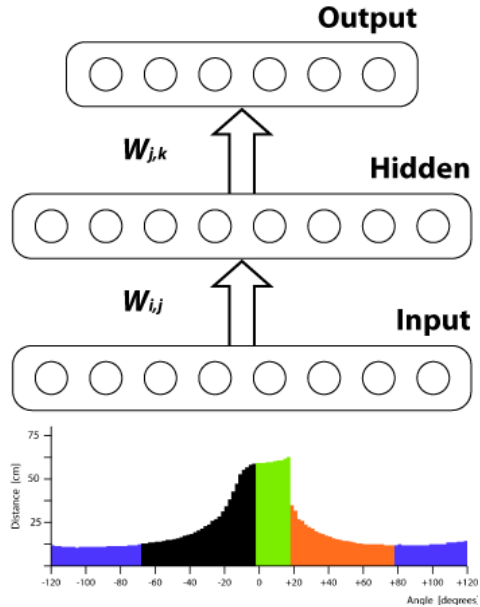

**Figure 3.** Representation of the multi-layer perceptron utilized to classify sensory patterns.

We have trained the neural network in such a way that, given a specific sensory pattern, in the output layer there will be only one highly active neuron which represents the best matching unit. Neurons in this layer project in a one-to-one fashion to the Hippocampus layer and to the Dorsal Striatum layer.

The second module is the Hippocampus and contains neurons that mimic place cells (green spheres in Figure 2). Each cell possesses a limited receptive field that covers only a small potion of the maze.

The number of effectively utilized units depends on the characteristics of the environment and on the order of the movements. It starts from zero and increases with exploration as unspecialized place cells learn to encode new parts of the environment.

One important characteristic of our architecture is that, unlike many existing models (Arleo and Gerstner, 2000; Dollé et al., 2010; Sheynikhovich et al., 2009), place cells in the Hippocampus are not hardwired to each other to form an graph-like representation of the maze. Instead, the transition from one place cell to the other is achieved by mentally simulating the movement from one location to the other.

Hippocampal cells are connected to neurons in the ventral striatum in an all-to-all fashion.

The third module represents the Striatum and has been subdivided into two subregions: ventral and dorsal. Neurons in the ventral striatum convey information about the value related to the connected place neurons. In this implementation, hippocampal and striatal neurons act in cooperation to provide spatial value information in the following way (see also (Chersi and Pezzulo, 2012; Pezzulo et al., 2013)). When the rat reaches a location in the maze, if a corresponding place cell exists it will start to fire at a high rate. This activation in turn elicits the firing of cells in the Ventral Striatum (due to the strong connectivity between the two areas) with an intensity that is proportional to the strength of their connections. In this view, the firing rate of a striatal neuron provides an indication of the proximity of the corresponding type of reward to the agent's position (and thus the probability of obtaining it). In addition to this, the input to the VS from the prefrontal cortex and the limbic areas can produce a modulation of the activity of striatal neurons (Corbit and Balleine, 2003). The projections from the Hippocampus and limbic areas to the Ventral Striatum allow to easily combine information about places and their potential value (in the sense of the reward the agent expects to receive when starting from that position) using a distributed representation.

The dorsal striatum is not directly connected to the hippocampus but instead to the sensory areas (Pennartz et al., 2011). Its neurons convey information related to the value of sensory stimuli. More precisely, in each position the rat sees an input patter described by equation 1. This pattern is processed and categorized by the sensory areas that provide as output the class to which the pattern belongs.

In the current implementation we have assigned 10 neurons to each reward type. This multiplicity is needed in our model to obtain the necessary variability in the collective neural responses (see below).

**Mathematical details**

In this implementation each unit of the network represents a small subpopulation of neurons encoding information specific to the area it belongs: locations in the hippocampus, value in the ventral striatum, and movements in the motor cortex.

The behavior of each neuronal pool is described by a firing rate model with synaptic currents (Dayan and Abbott 2001). This allows us to compactly represent complex interactions between excitatory and inhibitory neurons within pools and to explicitly take into account the dynamics of ionic currents and neurotransmitters. The set of equations governing the behavior of a single neuronal pool is the following:

$$\begin{cases} \tau_I \dfrac{dI_{syn}}{dt} = -I_{syn} + I_{ext} \\ v = g(I_{syn}) \\ \tau_J \dfrac{dJ}{dt} = -J + v \cdot (1-J) \end{cases} \quad (3)$$

where $I_{syn}$ is the total synaptic current, and $\tau_I = 25$ ms the corresponding time constant, $v$ is the firing rate, $J$ is a current due to slow neurotransmitters that is used as an eligibility trace for learning (see below), $\tau_J = 500$ ms is the corresponding time constant, $g()$ is the current-to-firing rate response function of a pool, which in this case has been modeled as:

$$\begin{cases} g(I) = g_{max} \cdot \tanh[\gamma(I - I_{thr})] & \text{for } I > I_{thr} \\ g(I) = 0 & \text{for } I < I_{thr} \end{cases} \quad (4)$$

where $g_{max}$ determines the maximum firing rate and $\gamma$ the steepness of the response function, and $I_{thr}$ is the firing threshold below which no response is present. In this implementation $g_{max} = 120$ Hz, $\gamma = 3 \cdot 10^9$ $A^{-1}$ and $I_{thr} = 4 \cdot 10^{-10}$ A. All the parameters have been chosen in order to obtain a biologically realistic model.

**Learning**

In this architecture there are three types of learning taking place within and between different areas of the network. In the following we will describe these separately for the sake of clarity, but in fact they occur more or less simultaneously.

On one side, plastic connections between hippocampal and striatal neurons are updated utilizing a dopamine modulated eligibility trace-dependent rule (Sutton and Barto, 1998): when the animal reaches a foraging site the VS neurons corresponding to the specific type of reward are activated and the network experiences a dopamine increase, which enables the learning mechanism according to the following equation:

$$\Delta w_{ik} = \lambda \cdot J_i \cdot v_k^{(VS)} \cdot DA + \varepsilon \quad (5)$$

where $w_{ik}$ is the weight between the *i-th* neurons in the hippocampus and the *k-th* neuron in the ventral striatum, $\lambda$ is the learning rate, $J_i$ is the slow current of the Hippocampal neuron, $v^{(VS)}_k$ is the firing rate of the *k-th* VS neuron, *DA* is the dopamine change. In order to avoid unbounded increase of the synaptic weights we have introduced a normalization mechanism that intervenes when $v^{(VS)}_k$ crosses the *100 spk/s* threshold.

Note that the slow current *J*, which does not contribute to the neuronal firing rate, is used here as an eligibility trace for modifying synaptic weights between the previously activated place cells earlier and striatal value neurons. Finally, ε is a small random value that is used to introduce variability in the connections' strength values. In addition to rendering the model biologically more realistic, this variability is fundamental in order to obtain a VS population of neurons that respond in a slightly different way. We point out the fact that as learning proceeds the variability of the VS neuronal responses due to ε decreases as all the HPC-VS connections receive on average the same amount of additional noise.

As training proceeds, place cells become more strongly connected to the VS neurons in a way that is directly proportional to their proximity to the goal place they encode (because the eligibility trace is more intense closer to the reward site). The final result of this learning procedure is the formation of neural maps with a weight gradient leading towards the foraging locations (see Figure 4).

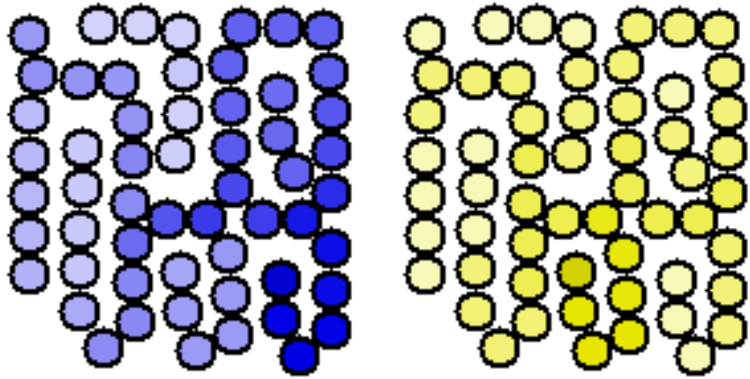

**Figure 4.** Connection strength between the place cells representing the maze in Figure 1 and the two striatal neurons encoding the different rewards (yellow = cheese, blue = water) after 30 trials. Darker colors mean stronger weights. Due to the employed learning rule these weights form a gradient that increases in direction of the reward locations.

Note that, VS neurons are not expressively liked to a specific position and the value they encode can change over time reflecting the occurrence location of the specific reward they code. In practical terms, if we suppose a reward is given in the right arm of a T-maze, the VS neuron coding that reward will start to build a gradient leading towards the rewarding site (see first two panel of Figure 5). If we then change the rewarding location to the end of the left arm, the same neuron(s) will remap their firing rate, gradually forming a gradient that points towards the new site (see last two panels in Figure 5).

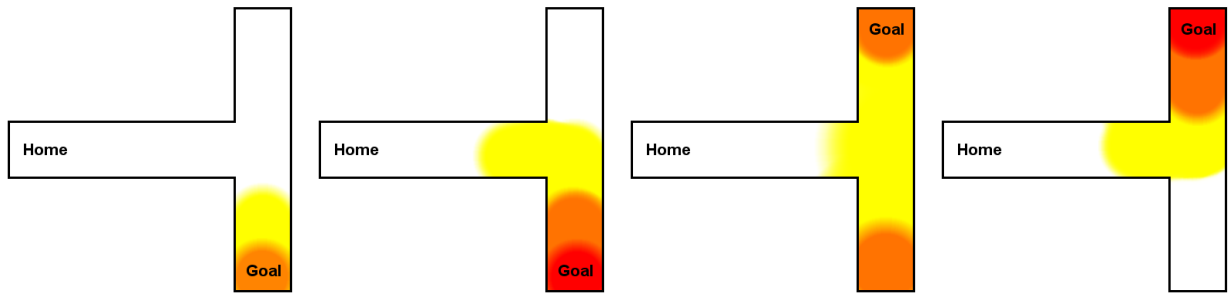

**Figure 5.** Various phases of the formation of the gradient in the representation of place value in a VS neuron. Darker colors mean higher firing rate.

The dorsal striatum employs a similar mechanism. More precisely, each place cell projects to the whole DS population which in our case selectively represents the value of all possible actions (i.e. turn left, turn right, go forward, go backward). When the rat walks in the maze, the executed actions produce an eligibility trace that is used to update the connections between the HPC and the DS according to eq. 5. The end result is a connection matrix that associates every place cell with the corresponding value of all possible actions in relation to a specific goal (see Figure 6). More specifically, the images below show the connection strength between the sensory states (mapped onto the maze) and the DS cells encoding "moving right", "moving down", "moving up" and "left" towards the blue goal (i.e. bottom right corner of the maze), respectively.

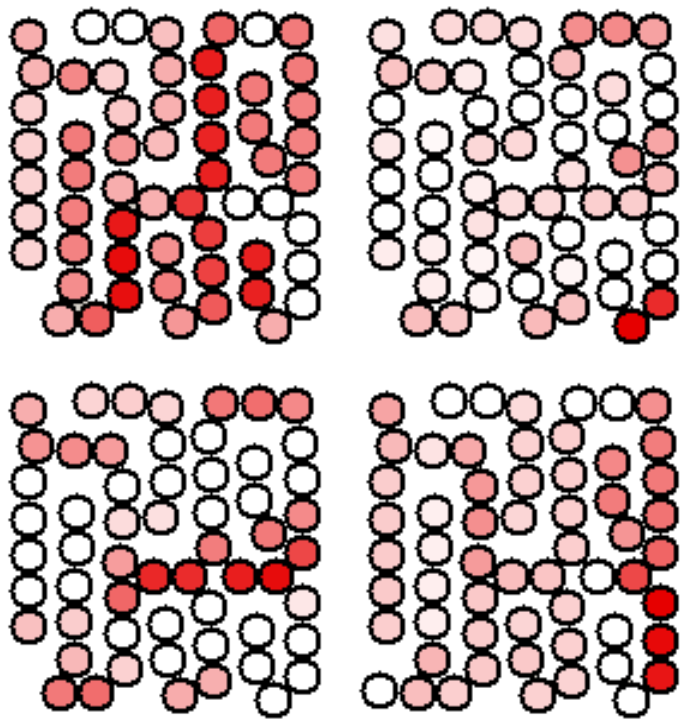

**Figure 6.** Representation of the connection strength between Hippocampal and Dorsal Striatum neurons. Each image represents the average strength of the connections with neurons encoding a specific movement direction: "right", "down", "up" and "left", respectively.

Finally, the multi-layer perceptron (i.e. the visual processing layer) is trained with a standard back-propagation algorithm. Nevertheless, in order to obtain a higher disambiguation capacity of the network we chose to add an extra term that penalizes the matching of a previously best matching unit when the animal moves away from the previous receptive field for more than a threshold value (in our case 2.5 cm).

**Landmarks**

One important aspect introduced in this model is the presence of landmarks (see figure 2). Physically, these are locations that possess special characteristics not present in other parts of the maze (in our case a red color patch). In our implementation, the presence of a landmark has the effect of multiplying the eligibility by a "saliency" factor causing the trace to be higher for these neurons. The result is that the value gradient, which is generally smoothly increasing toward to direction of the reward site, now presents local maxima in the points where landmarks are situated. These maxima are used during path evaluation to choose the optimal direction. Note that without these local maxima, in regions far away from the rewarding sites the value gradient would be completely flat, and thus uninformative.

Note that, once landmarks have been reached, they have to be ignored and a new (sub)goal has to be taken in consideration. In our model, this behavior has been obtained by hypothesizing a temporary suppression of the landmark's contribution to the global gradient once this has been reached. More precisely, in addition to neurons encoding the value of reward sites, the VS module also contains neurons encoding the value of landmarks (see Figure 7). When the rat reaches a landmark, the landmark gradient is temporarily subtracted from the global gradient (which by construction includes also all the landmarks' contributions). The result is a "corrected" gradient where the landmark is temporarily invisible thus allowing the rat to find the direction to the following landmark or goal. Without this temporary suppression, in the surroundings of a landmark there would always be a gradient pulling towards it causing the rat never to be able to leave that location.

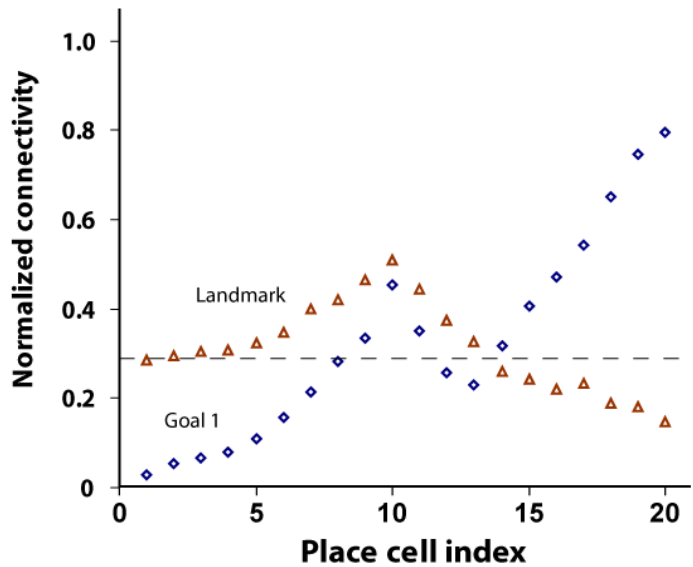

**Figure 7.** Connection strength to the VS neurons along the shortest path from the home position to the water reward site (see figure 2). Place cell number 10 encodes the position of the landmark thus it presents a local maximum of the firing rate.

**The decision-making mechanism**

As stated above, in this setup the rat receives a reward only if it reaches either of the two sites containing food or water. Thus, its motor controller has been realized in such a way to minimize the path to travel in order to reach the reward.
The path planning architecture is composed of three main parts (see Figure 8):
1) A *random choice* generator.
2) A *habitual response* mechanisms.
3) A *goal-directed decision* component.
The first mechanism has been introduced to guarantee that there is always at least a minimal exploratory behavior. This is useful not so much in the initial phases when the other modules produce mostly random and noisy responses, but rather in a later phase to handle unexpected events (such as a change in the environment) when the behavior has become inflexible through learning.

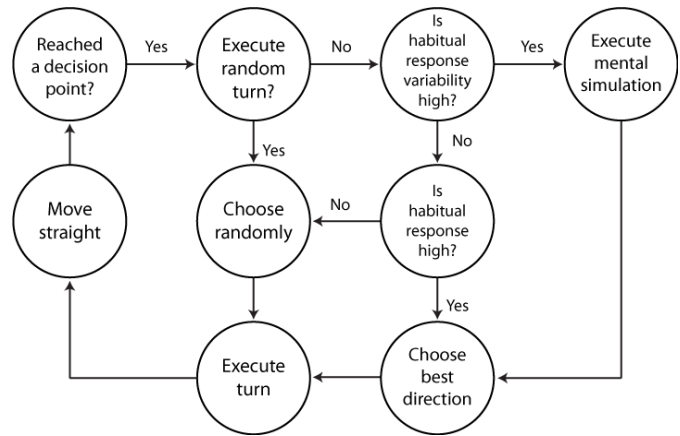

**Figure 8.** Schematic representation of the decision-making process.

The second mechanism utilizes the responses of neurons in the dorsal striatum. As described above, neurons in this area encode the value of actions with respect to a specific place, goal and context.
In this scenario, when the rat moves along the maze, the firing of place cells elicits the activation of DS neurons proportionally to the strength of their connections. Since the connectivity reflects the reward history obtained by executing the specific actions at each location, the firing rate level indicates the likeliness of reaching a desired goal if a certain action is executed. Following the scheme in figure 8, if the choice is not to be random, at each crossing point the average firing rate of all neurons for a given action (direction) is calculated as well as the associated variance. The choice is the result of the comparison between both the value (i.e. the firing rate) and its reliability of each direction.

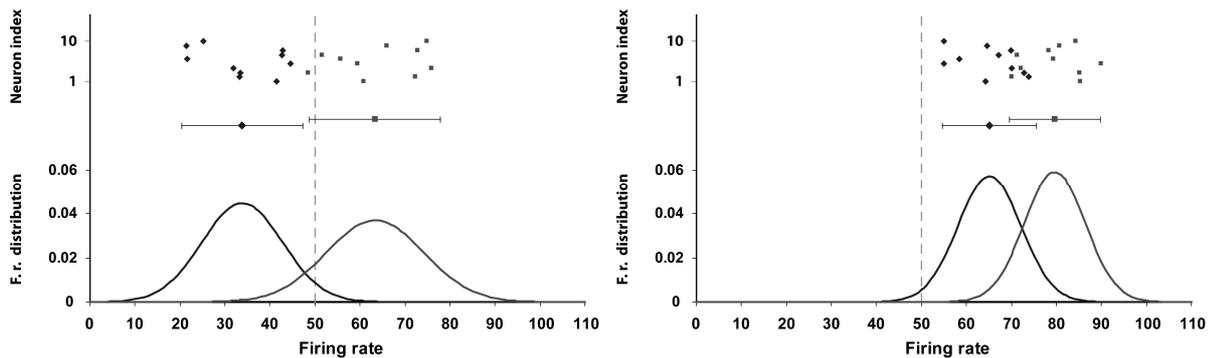

**Figure 9.** Two possible scenarios of the firing rate distributions of two subpopulations of neurons in the DS representing the "turn right" and "turn left" actions, respectively: 1) One direction (diamonds) has a mean activation level below significance threshold and one (squares) above but its variance is too large to be able to choose it with high confidence: mental simulation has to be executed to acquire more precise information. 2) Both directions are above significance level but their variance is too high to make an immediate choice so mental simulation is necessary to obtain more precise information.

Note that, as shown in Figure 8, there exits a hierarchy among the various modules, thus the second and third mechanism are not activated automatically at each step, instead they come into play only if the preceding mechanism provides unsatisfactory information. More precisely, at each decision point the system can choose (here with a probability of 15%) either a direction completely at random or to interrogate the habitual module, i.e. the dorsal striatum module, for the most valuable direction. As schematized in figure 9, the responses of the DS neural subpopulations can be combinations of the following types (being each response is independent):

1) The average activity of neurons coding one direction (e.g. left panel, blue squares) plus 1.5 sigmas is below the 50 spikes per second threshold: the value is considered not meaningful and random decision will to be taken.
2) The average activity is above threshold but subtracted by 1.5 sigmas it becomes non significant (left panel, red squares): mental simulation will be executed in order to determine a more precise value.
3) Both subpopulations are clearly above significance level, but their variances overlap so mental simulation is necessary to reduce their uncertainty.

The third mechanism exploits a model based approach to determine the best direction of motion. More precisely, starting from the current position the agent virtually projects itself first along one arm of the crossing and then along the other one. The predicted sensory states are fed to the multilayer network (see Figure 3), which processes the input and causes the activation of the corresponding neurons in the hippocampus. These neurons in turn project to VS neurons which provide the value of the position of interest. Repeating this procedure for a number of steps, allows the agent to retrieve (from memory) detailed information about the value (i.e. the possibility to reach a reward site) associated to each simulated path. ·

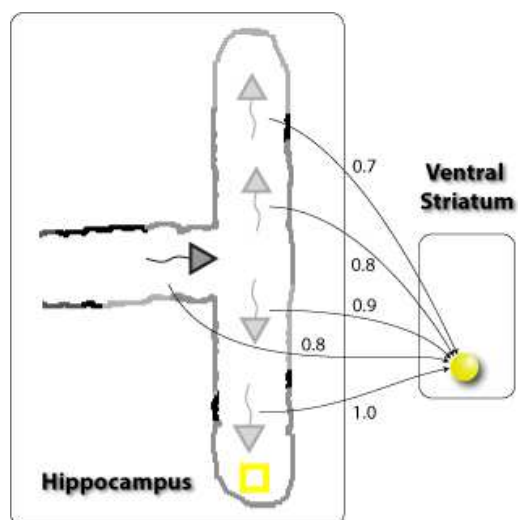

Figure 10. Schematic representation of the proposed decision mechanism based on mental simulation: at a T-crossing the rat imagines itself first moving along the right arm then along the left arm. Activity from hippocampal neurons

propagates to the Striatum eliciting a response that is proportional to the strength of the connections, thus providing a measure of the value of each choice.

The total value $U$ of a path is calculated as the average of all the activations $v_i$ of the striatal neuron representing the desired goal (food or water) along that path.

$$U = \frac{1}{N} \cdot \sum_{i=1}^{N} v_i \quad (8)$$

where $v_i$ is the response of the striatal neuron and $N$ is the number of simulated steps (where one step is equivalent to the size of the receptive field).

Decisions about the direction to take are initially completely random as all $U_i$ are very small, but as learning proceeds the neural responses become stronger and this begins to influence the choices favoring the direction of highest $U$ value.

**Conclusions and Discussion**

This paper proposes a biologically-inspired model of the core circuit involved in navigation and (spatial) decision making. It's main characteristics are presence of a habitual module, the sensory cortex - dorsal striatum axis, and a goal-directed module, the hippocampus – ventral striatum axis. Contrarily to existing models (Arleo and Gerstner, 2000; Dollé et al., 2010), here the hippocampus does not utilize reinforcement learning and does not form a graph-like representation of the environment on which it is possible to use methods like wave front propagation (Dollé et al., 2010; Ivey et al., 2011; Martinet et al., 2011; Niv et al., 2006). Instead, we have liked it to episodic memory, allowing it to rapidly learn associations between sensory states and place cells and to use goal-directed generative mechanisms, in our case mental simulation, to instantiate future states and estimate their value using the same machinery that is used during overt behavior.

Another difference with respect to existing models (Daw et al., 2005; Solway and Botvinick, 2012) is that here there is no direct competition between the two systems, but rather a support of the model-based to the model free mechanism: the habitual system automatically provides a fast (motor) response, but if this is too weak of too uncertain, the hippocampus comes into play to provide a slower but more precise response.

One of the important contributions of this paper is the detailed and biologically realistic simulation of how neurons in the different areas behave, what kind of information they code and how they interact with each other during navigation and spatial decision making. More specifically, we have shown how preplay of movement sequences (the "forward sweeps") may be generated in the brain

In this view, our results may be of great value for guiding the design and interpretation of behavioral and neurophysiological experiments.

Finally, we would like to point out the fact that the assumptions and the characteristics of this model are very general and valid not only for the navigation domain, but also for processes that involve memory, planning and future thinking.